\documentclass[times,doublespace]{simauth}

\usepackage{moreverb}

\usepackage[colorlinks,bookmarksopen,bookmarksnumbered,citecolor=red,urlcolor=red]{hyperref}

\newcommand\BibTeX{{\rmfamily B\kern-.05em \textsc{i\kern-.025em b}\kern-.08em
T\kern-.1667em\lower.7ex\hbox{E}\kern-.125emX}}

\def\volumeyear{2015}

\usepackage{lscape,rotating}
\usepackage{float,epsfig}
\usepackage{subfigure}
\usepackage{setspace}
 \usepackage{amsmath,bm} 
\usepackage{amsfonts}
\usepackage{amssymb}

\newtheorem{lemma}{Lemma}

\newenvironment{proof}[1][Proof]{\begin{trivlist}
\item[\hskip \labelsep {\bfseries #1}]}{\end{trivlist}}

\newcommand{\xv}{\bm{x}}
\newcommand{\yv}{\bm{y}}

\newcommand{\zv}{\bm{z}}

\newcommand{\muv}{\bm{\mu}}

\newcommand{\thetav}{\bm{\theta}}
\newcommand{\alphav}{\bm{\alpha}}
\newcommand{\betav}{\bm{\beta}}

\newcommand{\epsilonv}{\bm{\epsilon}}

\newcommand{\zerov}{\bm{0}}
\newcommand{\onev}{\bm{1}}

\makeatletter
\long\def\@makecaption#1#2{%
  \vskip\abovecaptionskip
  \sbox\@tempboxa{#1: #2}%
  \ifdim \wd\@tempboxa >\hsize
    #1: #2\par
  \else
    \global \@minipagefalse
    \hb@xt@\hsize{\box\@tempboxa\hfil}%
  \fi
  \vskip\belowcaptionskip}
\makeatother

\begin{document}

\runninghead{Y. Tang}

\title{Closed-form REML estimators and sample size determination for mixed effects models for repeated measures under monotone missingness}

\author{Yongqiang Tang}

\address{Shire, 300 Shire Way, Lexington, MA 02421, USA}

\corraddr{E-mail: yongqiang\_tang@yahoo.com}

\begin{abstract}
We derive the closed-form restricted maximum likelihood (REML) estimator and Kenward-Roger's variance estimator for 
 fixed effects  in the mixed effects model for repeated measures (MMRM) when the missing data pattern is monotone.  
As an important application of the analytic result, we present the formula for calculating the power of treatment comparison using the Wald t test with the Kenward-Roger adjusted variance estimate 
in MMRM. It allows adjustment for baseline covariates without the need to specify the covariate distribution in randomized trials.
A simple two-step procedure is proposed to determine the sample size needed to achieve the targeted power.
The proposed method performs well for both normal and moderately nonnormal data even  in small samples ($n=20$) in simulations. 
An anti-depressant trial is analyzed for illustrative purposes.
\end{abstract}


\maketitle

\section{Introduction}
\setlength{\parindent}{3ex}
In clinical trials,  the mixed effects model for repeated measures (MMRM) is commonly used to analyze   longitudinal continuous outcomes
collected at a number of fixed time points \cite{aiddiqui:2009, aiddiqui:2011, lu:2008}. 
In MMRM, 
there is no random effect, and the within-subject dependence is modeled by an unstructured covariance matrix.  It assumes that 
the post-baseline outcome $\yv_i =(y_{i1},\ldots,y_{ip})'$ follows a $p$-dimensional multivariate normal distribution
\begin{equation}\label{mixed1}
\yv_i \sim N_p[(\alphav_1' \xv_i,\ldots,\alphav_p' \xv_i)',\Sigma],
\end{equation} 
where $\xv_{i}$ is the baseline covariate (including intercept and treatment status) for subject $i$, and
$\alphav_j$ is the $q\times 1$ vector of covariate and treatment effects at visit $j$.
 In clinical trials, the primary analysis model is generally pre-specified in the protocol. 
While it is difficult, if not impossible, to correctly pre-specify the covariance structure, the use of an unstructured
 covariance matrix usually provides reasonable control of the type I error rate \cite{aiddiqui:2009}.
Any stronger assumptions on the mean response or covariance structure  can be difficult to be accepted by regulatory agencies without rigid justifications.

Some  theories for model \eqref{mixed1} can be presented in a simple and elegant way  \cite{little:2013} for monotone missing data. 
Closed-form maximum likelihood (ML) estimators \cite{anderson:1957} and Bayes estimators \cite{tang:2015} can be obtained 
by factoring the observed data distribution as $f(y_{i1},\ldots,y_{ij})=f(y_{i1}) \prod_{k=2}^j f(y_{ik}|y_{i1},\ldots,y_{ik-1})$.
These analytic results are  of great practical interest since the real clinical data  are often monotone or approximately monotone.
For near-monotone data,
monotone pattern can be obtained by removing subjects with intermittent missing data  \cite{little:2002}, by ignoring data collected after the first missing visit, 
or by
filling in the intermittent  missing data  using a single imputation method. A numerical example is provided in Section $2.4$. Alternatively, the intermittent missing data  can be handled
using the monotone expectation-maximization (MEM) algorithm \cite{liu:1999} or monotone data augmentation (MDA) algorithm 
\cite{tang:2015,2016:tang} whenever appropriate.
 
In Section $2$, we derive the closed-form restricted maximum likelihood (REML) estimator  and assess the biases
of ML and REML estimators for model \eqref{mixed1} under monotone missingness. 
We obtain the closed-form Kenward-Roger (KR) adjusted variance estimator  \cite{1997:kenward}  and compare it with the delta variance for inference on fixed effects. 
 An anti-depressant trial  \cite{2013:mallinckrodta} is analyzed for illustrative purposes.

Sample size calculation is critical in planning a clinical trial. If the sample size is too small, the trial may have little chance to detect a clinically meaningful treatment effect,
whereas an unnecessarily large sample may waste time, money and resources.
Sample size determination methods were previously developed for MMRM  \cite{lu:2008, lu:2009} based on the asymptotic variance of the treatment effect and/or normal approximation.
Since the use of the asymptotic variance underestimates the true variance of the estimated treatment effect \cite{1997:kenward}, and
the normal distribution can not adequately approximate the t distribution, the methods in  \cite{lu:2008, lu:2009} tend to  underestimate the required  size  especially
in small samples. 

In Section $3$, we derive the  power  formula for the KR adjusted Wald  t test comparing two treatments in MMRM. The power is computed based on the  KR adjusted variance 
and t test. Therefore, the proposed method tends to be more accurate than the methods of \cite{lu:2008, lu:2009}, but the corresponding formulae have simpler expressions.
One advantage of our method is that 
it allows adjustment for covariates, but there is no need to specify the covariate distribution. 
Information about the covariate distribution is usually not available at the design stage since 
subjects will be enrolled into the trials only if they meet certain inclusion/exclusion criteria.
There is no closed-form solution for the sample size  needed to achieve a targeted power.
We propose a  two-step procedure to approximate the required  size and assess its performance  for both normal and non-normal continuous  data in small to moderate samples via simulation. 

We focus on the monotone missing data. Section $5$ will discuss how to apply the sample size calculation method to non-monotone missing data.

\section{REML estimator under monotone missingness}
\subsection{Closed-form REML estimator}
We assume the  missing data pattern is monotone in the sense that
  if $y_{ij}$ is observed, then $y_{it}$'s are observed for all $t\leq j$.
 Let $r_i$ denote the dropout pattern according to   the last observation. A subject is in pattern
$r_i=s$ if
$y_{ij}$ is observed for $j\leq s$,
 $r_i=0$ if $\yv_i$ is missing, and  $r_i=p$   if the subject  completes all $p$ visits.  Without loss of generality, we sort the data so that subjects in
pattern $s$ are  before subjects in pattern $t$ if $s>t$.  For notational simplicity, we assume there are  two treatment groups ($g_i=1$ for active treatment, $g_i=0$ for placebo). 
Let $\mathcal{B}_{gs}$ denote the subset of subjects retained at visit $s$ ($y_{is}$ is observed) in  group $g$, 
 $n_{gs}$ the number of subjects in $\mathcal{B}_{gs}$, and  $n_s=\sum_{g=0}^1 n_{gs}$.
Let $\xv_{bi}$ denote the baseline covariates excluding the intercept term and $g_i$. The baseline outcome  $y_{i0}$ is usually included as a covariate in $\xv_{bi}$.
Let $\bar\xv_{{gs}}$ be the average of $\xv_{bi}$ among subjects in $\mathcal{B}_{gs}$.

The clinical data are often analyzed  by model \eqref{mixed1}, where  $\xv_i =(1,\xv_{bi}', g_i)'$ and $\alphav_j=(\mu_j,\alphav_{jb}',\tau_j)'$.  
 The ML estimator of the variance parameters 
is  biased downward since it does not take into account of the loss in degrees of freedom (d.f.) from the estimation of  $\alphav_j$'s \cite{harville:1977}.
The REML procedure is generally employed to reduce the bias in estimating the variance parameters, and inference for fixed effects
 is often made based on the KR   variance estimate especially in small samples. 
Let $W_{ij}$ be a $pq\times 1$ vector whose $[(j-1)q+1]$-th through $[jq]$-th entries are given by $\xv_i$ while other entries are  $0$'s.
Let $W_i=(W_{i1},\ldots,W_{ir_i})'$ and $\yv_{io}=(y_{i1},\ldots,y_{ir_i})'$. Let $W=(W_1',\ldots,W_n')'$ and $Y=(\yv_{1o}',\ldots,\yv_{no}')'$. The restricted likelihood
is the marginal likelihood of $KY$, where $N=\sum_{i=1}^n r_i$ and $K$ is any $(N-pq) \times N$ matrix of full rank satisfying $KW=\zerov$.
The restricted log-likelihood $\ell_r$ does not depend on $\alphav$ and the  choice of $K$, and can be expressed as \cite{harville:1974, VERBYLA:1990}
\begin{eqnarray}\label{reml1}
\begin{aligned}
 2\ell_r   =& -\sum_{i=1}^n\log{|\Sigma_{r_i}|} -\log{|\sum_{i=1}^n W_i'\Sigma_{r_i}^{-1}W_i|} -\left[\sum_{i=1}^n \yv_{io}'\Sigma_{r_i}^{-1}\yv_{io}  \right. \\
  &\,\qquad\qquad \left.-(\sum_{i=1}^n \yv_{io}'\Sigma_{r_i}^{-1}W_{i})(\sum_{i=1}^n W_i'\Sigma_{r_i}^{-1}W_i)^{-1} (\sum_{i=1}^n W_{i}'\Sigma_{r_i}^{-1} \yv_{io})\right], 
\end{aligned}
\end{eqnarray}
where $\Sigma_k$ is the leading  $k\times k$ submatrix of $\Sigma$.

Let $\Sigma=L\Lambda L'$ be the LDL decomposition of $\Sigma$, where $U=\begin{bmatrix} 1 & 0 &  \ldots & 0\\
                       -\beta_{21} & 1 & \ldots &0 \\
                                       & \ldots & \ldots & 0\\
                      -\beta_{p1} & \ldots & -\beta_{p,p-1} & 1 \\
                    \end{bmatrix}$,   $L=U^{-1}$ and $\Lambda=\text{diag}(\sigma_1^2,\ldots,\sigma_p^2)$.  Let $l_{jt}$ be the $(j,t)$ entry of $L$.  
The decomposition admits nice interpretation. 
The entries $\betav_j=(\beta_{j1},\ldots,\beta_{jj-1})'$ are the regression parameters of $y_{ij}$ on 
$\vec{\yv}_{j-1}=(y_{i1},\ldots,y_{i,j-1})'$  since model  \eqref{mixed1} can be written as 
\begin{equation}\label{factor0}
 (\underline{y}_{i1},\ldots,\underline{y}_{ip})' \sim N_p[(\underline{\alphav}_1'\xv_i, \ldots,\underline{\alphav}_p'\xv_i)', \Lambda],
\end{equation}
which can be reorganized as \cite{2016:tang}
\begin{equation}\label{factor} 
y_{ij}= \xv_i'\underline{\alphav}_j +\vec\yv_{ij-1}'\betav_j +\varepsilon_{ij}=\zv_{i,j-1}'\thetav_j+\varepsilon_{ij} \text{ for } j\leq p,
\end{equation}
 where  $\underline{y}_{ij} = y_{ij} - \sum_{t=1}^{j-1} \beta_{jt} y_{it} $, $\underline{\alphav}_{j} =\alphav_{j} - \sum_{t=1}^{j-1} \beta_{jt} \alphav_{t} $, 
 $\thetav_j=(\underline{\alphav}_{j}', \betav_j')'$, 
    $\zv_{ij} =(\xv_i',\vec{\yv}_{ij}')'$,  $\varepsilon_{ij} \sim N(0,\sigma_j^2)$, and $\varepsilon_{ij}$'s are independent. 

Let $X_{o_j}=[\xv_{1},\ldots,\xv_{n_j}]'$, $Z_{o_j}=[\zv_{1j},\ldots,\zv_{n_jj}]'$, and $Y_{o_j}=(y_{1j},\ldots,y_{n_jj})'$.
The least square (LS) estimator for the parameters in model \eqref{factor} is 
\begin{equation*}\label{lsest}
\hat\thetav_j =(Z_{o_j}'Z_{o_j})^{-1}Z_{o_j}'Y_{o_j} \text{ and } \hat\sigma_j^2=\hat{S}_j/(n_j-q-j+1),
\end{equation*}
 where 
$\hat{S}_j =\sum_{i=1}^{n_j}(y_{ij}- \zv_{ij-1}'\hat\thetav_j)^2$.  The ML estimator for the parameters  in model \eqref{factor} is  given by
\begin{equation*}\label{mlest}
\hat\thetav_{j,ml}=\hat\thetav_j \text{ and } \hat\sigma_{j,ml}^2= \hat{S}_j/n_j.
\end{equation*}

Lemma $1$  derives the closed-form REML estimator for the parameters in model \eqref{mixed1}. Its proof is given in the appendix.
\begin{lemma}\label{solreml}
(a) The asymptotic variance of the fixed effect estimate $(\hat\alphav_1',\ldots,\hat\alphav_p')'$ is
$(\sum_{i=1}^n  W_i'\Sigma_{r_i}^{-1}W_i)^{-1} =(L\otimes I_q) V_w (L'\otimes I_q)$, where $I_q$ is the $q\times q$ identity matrix,
and $V_w =\text{\normalfont diag}[\sigma_1^2(X_{o_1}'X_{o_1})^{-1},\ldots, \sigma_p^2(X_{o_p}'X_{o_p})^{-1}]$.\\
(b) The restricted log-likelihood \eqref{reml1} can be decomposed as the sum of $p$ independent log-likelihoods $\ell_r=\sum_{j=1}^p \ell_{r_j}+\text{constant}$, where
\begin{equation}\label{reml2}
\ell_{r_j} =\frac{q-n_j}{2}\log(\sigma_j^2)-\frac{(Y_{o_j}-\vec{Y}_{o_j}\betav_j)'Q_{j}(Y_{o_j}-\vec{Y}_{o_j}\betav_j)}{2\sigma_j^{2}},
\end{equation}
$Q_j=I-X_{o_j}(X_{o_j}'X_{o_j})^{-1}X_{o_j}$, and
$\vec{Y}_{o_j}$ is a $n_j \times (j-1)$ matrix whose $(i,t)$ entry is $y_{it}$.\\
(c) The REML  estimator is $\hat\betav_j=( \vec{Y}_{o_j}'Q_j\vec{Y}_{o_j})^{-1}\vec{Y}_{o_j}'Q_jY_{o_j}$,  $\underline{\hat\alphav}_j=(X_{o_j}'X_{o_j})^{-1}X_{o_j}'(Y_{o_j}-\vec{Y}_{o_j}\hat\betav_j)$
and $\hat\sigma_{j,re}^2=\hat{S}_j/(n_j-q)$. Furthermore, $\hat\theta_{j,re} =(\underline{\hat\alphav}_j', \hat\betav_j')'=\hat\thetav_j$. \\
(d) The asymptotic variance of $\hat\betav_j$ based on the observed or expected information matrix is given respectively by 
$$ \text{\normalfont var}_{_O}(\hat\betav_j) =\sigma_j^2( \vec{Y}_{o_j}'Q_j\vec{Y}_{o_j})^{-1} 
\text{ and } \text{\normalfont var}_{_E}(\hat\betav_j) =\sigma_j^2[\text{\normalfont E}(\vec{Y}_{o_j}'Q_j\vec{Y}_{o_j})]^{-1}= \frac{\sigma_j^2}{n_j-q}\Sigma_{j-1}^{-1}.$$
\end{lemma}

{\flushleft Remarks:}\\
 1) Lemma 1(a) is also valid for models with structured covariance matrix; \\
2) We set $\vec{Y}_{o_j}\betav_j=0$ at $j=1$ since $\betav_j$ ($\vec{Y}_{o_j}$) is an empty vector (matrix);\\
3)  $(\vec{Y}_{o_j}'Q_j\vec{Y}_{o_j})^{-1}$ is identical to the lower-right $(j-1)\times (j-1)$ submatrix of $(Z_{o_j}'Z_{o_j})^{-1}$.

Lemma 1(b) can also be derived from the Bayesian viewpoint based on the fact \cite{harville:1974}
that  the restricted likelihood is proportional to the marginal posterior  density over the variance parameters under a flat prior for $(\alpha,\Sigma)$.
We have to ignore the Jacobian correction factor in the posterior density under reparameterization since the REML estimators are invariant under reparameterization. 
Under a flat prior, $(\thetav_j,\sigma_j^2)$'s are independent in the posterior distribution, and the posterior density of $(\thetav_j,\sigma_j^2)$ is proportional to the  likelihood function for model 
\eqref{factor}, which implies that
\begin{eqnarray*}
\begin{aligned}
 \sigma_j^2| Y & \sim \hat{S}_j/\chi_{n_j-q-j-1}^2, \\
\betav_j|\sigma_j^2,Y & \sim N_{j-1}(\hat\betav_j, \sigma_j^2 [\vec{Y}_{o_j}'Q_j\vec{Y}_{o_j}]^{-1}), \\
\alphav_j|\sigma_j^2,\betav_j, Y & \sim N_q[(X_{o_j}'X_{o_j})^{-1}X_{o_j}'(Y_{o_j}-\vec{Y}_{o_j}\betav_j), \sigma_j^2(X_{o_j}'X_{o_j})^{-1}].
\end{aligned}
\end{eqnarray*}
The marginal posterior distribution of $(\betav_j,\sigma_j^2)$ is given by
 $$f(\sigma_j^2,\betav_j|Y)=f(\sigma_j^2|Y)f(\betav_j|\sigma_j^2,Y)\propto \exp(\ell_{r_j}).$$

\subsection{Bias in parameter  estimates}
The LS, ML and REML methods produce identical estimate  of $\thetav_j$'s. Since $\hat\thetav_j$ is an unbiased estimate of $\thetav_j$, it is easy to show
by induction that  $(\hat{l}_{j1},\ldots,\hat{l}_{jj})$ and $\hat\alphav_j= \sum_{t=1}^j \hat{l}_{jt} \hat{\underline\alphav}_t$ are  unbiased respectively for $({l}_{j1},\ldots,{l}_{jj})$ and  $\alphav_j= \sum_{t=1}^j l_{jt} \underline\alphav_t$.

 For $\sigma_j^2$, the LS estimator is unbiased. The bias $\text{E}(\hat\sigma_{j,re}^2) -\sigma_j^2= - (j-1)\sigma_j^2/(n_j-q)$ of the REML estimator is smaller than that
$\text{E}(\hat\sigma_{j,ml}^2) -\sigma_j^2= - (q+j-1)\sigma_j^2/n_j$ of the ML estimator.

The bias is not invariant to reparameterization. In REML, the bias in $\hat\Sigma = \hat{L}\hat\Lambda\hat{L}'$ is of the order of $O(n^{-2})$.
It can be derived by the Cox-Snell \cite{cox:1968} method (using the equation on line $21$, page $2586$ of \cite{kenward:2009} and
 equation (9.62) of \cite{pase:1997}), or 
based on the second-order Taylor series expansion of $\hat{\Sigma}-\Sigma= \hat{L}\hat\Lambda\hat{L}'- L\Lambda\,L' $ around 
$(\betav_j,\sigma_j^2)$'s. In ML, the bias in $\hat\Sigma_{\text{ML}}$ is $O(n^{-1})$ since
 $\text{E}(\hat\Sigma_{\text{re}}-\hat\Sigma_{\text{ML}})  = \text{E}[\hat{L}\, \text{diag}(\varphi_1,\ldots,\varphi_p)\hat{L}']$,
where $\varphi_j=\text{E}(\hat\sigma_{j,re}^2-\hat\sigma_{j,ml}^2 )= O(n^{-1})$. The bias   in the LS estimator of $\Sigma$ is also $O(n^{-1})$.

\subsection{Variance of fixed effects}
The fixed effect estimate is $\hat\alphav= (\hat\alphav_1,\ldots, \hat\alphav_p)' = \hat{L} ( \hat{\underline\alphav}_1, \ldots, \hat{\underline\alphav}_p)'$,
 where $\hat{\underline\alphav}_j$  is defined in Lemma \ref{solreml}c.
 The  asymptotic variance of $\hat\alphav$  given in Lemma \ref{solreml}a
 is in fact the variance of  the  estimate $(\tilde\alphav_1',\ldots, \tilde\alphav_p')'$ assuming that the variance parameters $(\betav_j,\sigma_j^2)$'s and hence  $L$ are known,
where  $(\tilde\alphav_1,\ldots, \tilde\alphav_p)' = L ( \tilde{\underline\alphav}_1, \ldots, \tilde{\underline\alphav}_p)'$,  $\tilde{\underline\alphav}_j= (X_{o_j}'X_{o_j})^{-1}X_{o_j}'(Y_{o_j}-\vec{Y}_{o_j}\betav_j)$,
and $\text{var}(\tilde{\underline\alphav}_j)=\sigma_j^2 (X_{o_j}'X_{o_j})^{-1}$. Let $\Phi_p=\text{var}(\tilde\alphav_p)=\sum_{j=1}^p l_{pj}^2\sigma_j^2 (X_{o_j}'X_{o_j})^{-1}$.

 In the appendix, we show $\text{cov}(\hat\alphav_p-\tilde\alphav_p,\tilde\alphav_p)=\zerov$, 
and derive the variance of $\hat\alphav_p-\tilde\alphav_p$ 
\begin{equation}\label{varalphadiff}
\Psi_p=\text{\normalfont var}(\hat\alphav_p-\tilde\alphav_p) = \sum_{j=2}^p l_{pj}^2 (V_{dj}+V_{ej}) \approx  \sum_{j=2}^p l_{pj}^2 V_{dj},
\end{equation}
where $L_j$ is the leading $j\times j$ submatrix of $L$, $(\omega_{j1},\ldots,\omega_{j,j-1})$ are the diagonal elements of $L_{j-1}' \text{\normalfont var}(\hat\betav_j)  L_{j-1}$ for $j\geq 2$,
 $D_{jt}=  ( X_{o_j}'  X_{o_j})^{-1}-(X_{o_t}'X_{o_t})^{-1}$,  $V_{dj}= \sum_{t=1}^{j-1} \omega_{jt} \sigma_t^2 D_{jt}$,  $V_{e1}=\zerov$, 
$V_{ej}= \sum_{t=1}^{j-1} \omega_{jt} (V_{dt}+V_{et})$ at $j\geq 2$, $V_{ej}$ is of lower order than $V_{dj}$ and can  be ignored. Thus
\begin{equation}\label{varalpha} 
\text{\normalfont var}(\hat\alphav_p) =\Phi_p + \Psi_p=\sum_{j=1}^p l_{pj}^2\sigma_j^2 (X_{o_j}'X_{o_j})^{-1}+\sum_{j=2}^p l_{pj}^2 V_{dj}.
\end{equation}
Although $\Psi_p$  is of lower order than $\Phi_p$, it is  not negligible in small samples.
The variance at other visits can be derived similarly, and would not be presented here.

Let $\Delta_{s}=\bar{\xv}_{{1s}}-\bar{\xv}_{{0s}}$, $S_{x_{s}}=\sum_{g=0}^1 \sum_{i\in\mathcal{B}_{gs}} (\xv_{bi}-\bar{\xv}_{{gs}})^{\otimes 2}$ and
 $V_{x_s}= n_{1s}^{-1}+n_{0s}^{-1}+\Delta_{s}' S_{x_{s}}^{-1}\Delta_{s}$, 
where $a^{\otimes 2}=aa'$. By equation \eqref{varalpha}, the  variance of the treatment effect estimate $\hat\tau_p$ is given by
\begin{equation}\label{vartau}
\text{var}(\hat\tau_p)= \sum_{j=1}^p l_{pj}^2\sigma_j^2  V_{x_j} + \sum_{j=2}^p l_{pj}^2  \sum_{t=1}^{j-1} \omega_{jt} \sigma_t^2 (V_{x_j}-V_{x_t}).
\end{equation}

\subsubsection{Kenward-Roger (KR)  variance estimate} The KR variance estimate has been widely used in practice particularly when the sample size is small. The KR approach 
not only takes into account of 
the variability in the variance parameter estimate, but also  adjusts for the bias in $\hat\Phi_p$. In the appendix, we show 
the bias in $\hat\Phi_p$ is  
\begin{eqnarray}\label{biasasy}
\begin{aligned}
\Psi_p^*=\text{E}(\hat\Phi_p)-\Phi_p=&\sum_{j=2}^p  \sum_{t=1}^{j-1}  l_{pj}^2\omega_{jt} \sigma_t^2   (X_{o_t}'X_{o_t})^{-1}  -  \sum_{j=2}^p  \frac{j-1}{n_j-q}\sigma_j^2 l_{pj}^2  (X_{o_j}'X_{o_j})^{-1}.
\end{aligned}
\end{eqnarray} 
If $\text{\normalfont var}_{_E}(\hat\betav_j)$  in Lemma \ref{solreml}d is used, then 
$\omega_{jt}\sigma_t^2 = \sigma_j^2/(n_j-q)$,
$\Psi_p=\sum_{j=2}^{p} l_{pj}^2 \sigma_j^2 [\sum_{t=1}^{j-1} D_{jt}]/(n_j-q)$,
and $\Psi_p^*=-\Psi_p$.
The KR variance  evaluated at the REML estimator  is given by
\begin{equation}\label{varkr}
\widehat{\textrm{var}}_{\text{kr}}(\hat\alphav_p)=\hat\Phi_p+\hat\Psi_p -\hat\Psi_p^* =\hat\Phi_p+2\hat\Psi_p,
\end{equation}
and it provides a roughly unbiased estimate of the variance 
$\Phi_p+\Psi_p$ of $\hat\alphav_p$ while the lower order terms are dropped.  The KR 
variance estimate is invariant under reparameterization of an unstructured covariance matrix \cite{kenward:2009}.
 Although the derivation of the KR variance estimate relies on $\text{\normalfont var}_{_E}(\hat\betav_j)$, 
equation \eqref{varkr} yields the same result as Kenward-Roger's \cite{1997:kenward} formula if the  calculation is based on $\text{\normalfont var}_{_O}(\hat\betav_j)$.

\subsubsection{Delta variance estimate} 
An alternative variance estimate for  $\hat\alphav_p$ can be obtained from the delta method
$$V_{_\text{d}}  =  \sum_{j=1}^p \frac{\partial{\alphav_p}}{\partial\thetav_j} \text{var}(\hat{\thetav}_j)  (\frac{\partial{\alphav_p}}{\partial\thetav_j})'     = \sum_{j=1}^p l_{pj}^2  \sigma_j^2 \ddot\alphav_j (Z_{o_j}'Z_{o_j})^{-1} \ddot\alphav_j', $$
where $\partial{\alphav_p}/{\partial\thetav_{j}}= l_{pj} \ddot\alphav_j$ and $\ddot\alphav_j=(I_q,\alphav_1,\ldots,\alphav_{j-1})$.

By equation \eqref{matrixinv} in the appendix, the delta variance evaluated at the LS estimator can be written as
$$\hat{V}_{_\text{d}}=\sum_{j=1}^p \hat{l}_{pj}^2\hat\sigma_j^2 (X_{o_j}'X_{o_j})^{-1}+
         \sum_{j=2}^p \hat{ l}_{pj}^2  (\hat{\vec\alphav}_j-\hat{\vec\alphav}_j^*) V_{\betav_j} (\hat{\vec\alphav}_j-\hat{\vec\alphav}_j^*)',$$
where $\hat{\vec\alphav}_j=(\hat\alphav_1,\ldots,\hat\alphav_{j-1})$ and $\hat{\vec\alphav}_j^*=(X_{o_j}'X_{o_j})^{-1} X_{o_j}'\vec{Y}_{o_j}$.
The delta variance estimate tends to be more conservative  than the KR variance estimate since the 
 latter is a roughly unbiased variance estimate for $\hat\alphav_p$, while the former overestimates the variance of $\hat\alphav_p$ by 
\begin{eqnarray*}
\begin{aligned}
 \text{E}(\hat{V}_{_\text{d}}) -\Phi_p - \Psi_p 
  \approx \sum_{j=1}^p \text{var}(\hat{l}_{pj}) \sigma_j^2 (X_{o_j}'X_{o_j})^{-1} + \sum_{j=2}^p \text{var}(\hat{l}_{pj})   [\sum_{t=1}^{j-1} D_{jt} \omega_{jt}\sigma_t^2].
\end{aligned}
\end{eqnarray*}

The delta variance is closely related to the variance from the multiple imputation (MI) inference \cite{2016:tangc}.
One may compare the KR variance estimate given in equation \eqref{varkr} with   the MI variance derived by \cite{2016:tangc}.
In general, the MI inference tends to be slightly more conservative than the REML inference \cite{aiddiqui:2011}.

\subsection{Analysis of an antidepressant trial}\label{exam1}
To illustrate the analytic result derived in this section,  we analyze  an antidepressant trial reported by \cite{2013:mallinckrodta}. 
The Hamilton 17-item rating scale for depression ($\text{HAMD}_{17}$)  is collected at baseline 
and weeks 1, 2, 4, 6. The dataset consists of $84$ subjects on the active treatment, and $88$ subjects on placebo.
 The number of subjects in patterns $0$ to $4$ is respectively $(0, 7, 5, 11, 65)$ in the placebo arm, and  
$(0, 6, 5, 9, 64)$ in the active arm.
The missing data are mainly due to dropout. Only one subject has an intermittent missing value at visit $2$. 
We impute the intermittent missing value as  $11.70432$ based on the linear regression
 of $y_{i2}$ on $(y_{i0}, y_{i1}, g_i)$, where $y_{i0}$ is  the baseline $\text{HAMD}_{17}$.
A better approach would be to replace the missing value with multiple plausible values to reflect uncertainty about the missing data via the MI  procedure \cite{2016:tangb, 2016:tang}. 
However, due to the small amount of non-monotone missing data, the result from this single imputation approach is very close to that from the MI inference \cite{2016:tangb}.

Table \ref{varcomp} reports the result from the MMRM analysis.  The model includes visit, baseline $\times$ visit and treatment $\times$ visit  interactions as the fixed effects, and
an unstructured covariance matrix is used to model the within-patient errors. This corresponds to $\xv_i=(1, y_{i0}, g_i)'$ in terms of model \eqref{mixed1}. 
SAS PROC MIXED yields the same result as our analytic formulae if  a more stringent convergence criterion (CONVG= $1E$-$10$ option) than the default is used.
The KR variance estimate is  smaller than the delta variance, but slightly larger than the asymptotic variance. 
The $p$-value for testing the treatment effect at week $6$ is $0.0132$ based on the KR variance estimate.

Table \ref{varcomp} also reports the result from the constrained longitudinal data analysis (cLDA), in which the baseline outcome is treated as a response variable instead of a covariate, and 
 constrained to have equal mean  across treatment groups  \cite{liu:2009,lu:2010}. For monotone missing pattern, the 
REML estimate and the associated variance estimate can be derived using essentially the same technique described in Sections $2.1$ and $2.3$, and the details are omitted here.
 The REML estimates are in fact identical,  and the corresponding KR variance estimates are very close  in MMRM and cLDA. This is consistent with the simulation result reported in \cite{lu:2010}.
The cLDA is useful if there is  a non-negligible amount of missing data at baseline.
 However, for monotone missing data, the MMRM is generally preferred since there is no need to assume that $y_{i0}$'s are normally distributed
and that the baseline means  are equal across treatment groups in MMRM.

\begin{table}
\begin{center}
\caption{Treatment effect estimate and its standard error at week $6$ in an antidepressant trial
}\label{varcomp}
\begin{tabular}{lcccc} \\\hline 
           & Treatment &  \multicolumn{3}{c}{Standard Error} \\     \cline{3-5}
Method & effect &  Asymptotic & KR  & Delta \\ \hline
MMRM  &  $-2.799$ &   $1.114$  & $1.116$ & $1.122$    \\
cLDA    &  $-2.799$  &   $ 1.098$ & $ 1.108$ & -  \\
\hline
\end{tabular} 
\end{center}
\end{table}

\section{Power and sample size formulae for MMRM}
\subsection{Power and sample size formulae}
Throughout Section $3$, we assume the hypothesis of interest is to test the treatment effect at the last visit. 
Let $\gamma_g$ be the proportion of subjects randomized to  group $g$, 
 $\pi_{gt}$ the retention rate at visit $t$  in group $g$, and $\bar\pi_t=\sum_{g=0}^1 \gamma_g\pi_{gt}$  the pooled retention rate at  visit $t$.
Then  $n_{gt}= n\gamma_g \pi_{gt}$.
In the variance  expression \eqref{vartau}, $\xv_{bi}$'s are assumed to be fixed in the analysis, but unknown at the design stage. We  will replace $V_{x_j}$'s in equation \eqref{vartau} by their expected values.
 Suppose the covariates are balanced between two arms. We don't require the covariates to be continuous, or follow a specific distribution in randomized trials.
Let $q^*$ be the dimension of $\xv_{bi}$, $q=q^*+2$ 
and  $\varpi_{\pi_t}= \sum_{g=0}^1 (\gamma_g\pi_{gt})^{-1}$.
It is easy to show that $n_{0t}n_{1t}(n_t-q^*-1)/(n_tq^*) \Delta_{t}' S_{x_{t}}^{-1}\Delta_{t}$ approximately follows
a F distribution with $q^*$ and $n_t-q^*-1$ d.f.
 (this holds exactly if $\xv_{bi}$ is normally distributed). We have
\begin{equation}\label{expvar1}
\text{E}(\Delta_{t}' S_{x_{t}}^{-1}\Delta_{t})\approx \frac{\varpi_{\pi_t}}{n} \varpi_{\delta_t}(n) \text{ and } \text{E}(V_{x_t})\approx \frac{\varpi_{x_t}(n)}{n}.
\end{equation} 
where  $\varpi_{\delta_t}(n)=q^*/(n\bar\pi_t-q-1)$ and $\varpi_{x_t}(n)= \varpi_{\pi_t}[1+ \varpi_{\delta_t}(n)]$
are  functions of $n$.

If we use $\text{var}_{{_E}}(\betav_j)$, then $ \omega_{jt} \sigma_t^2 =\sigma_j^2/(n_j-q)$.
 The (expected) variance of $\hat\tau_p$ is $\varpi_{\tau}(n)/n$, where
$$\varpi_{\tau}(n)=\sum_{j=1}^p l_{pj}^2\sigma_j^2  \varpi_{x_j}(n) + \sum_{j=2}^p l_{pj}^2 \sigma_j^2 \frac{\sum_{t=1}^{j-1}(\varpi_{x_j}(n)-\varpi_{x_t}(n))}{n\bar\pi_j-q}.$$
The power of the Wald test at  a two-sided  significance level of $\alpha$ is given by 
\begin{eqnarray}\label{power0}
\begin{aligned}
P=\text{Pr}(|\frac{\hat{\tau}_p }{\sqrt{\widehat{\textrm{var}}_{\text{kr}}(\hat\tau_p)}}| \geq t_{f,1-\frac{\alpha}{2}})  =&
               \text{Pr}(t(f,\sqrt{n}\lambda) \geq  t_{f,1-\frac{\alpha}{2}} )+ \\
 & \qquad \text{Pr}(t(f,\sqrt{n}\lambda) \leq -t_{f,1-\frac{\alpha}{2}} )  \approx \text{Pr}( t(f,\sqrt{n}|\lambda|) \geq  t_{f,1-\frac{\alpha}{2}}) ,
\end{aligned}
\end{eqnarray}
where $\widehat{\textrm{var}}_{\text{kr}}(\hat\tau_p)$ is the $(q,q)$ entry of  $\widehat{\textrm{var}}_{\text{kr}}(\hat\alphav_p)$ defined in equation \eqref{varkr},  $\lambda=\tau_p/\sqrt{ \varpi_\tau(n)}$,
 $t(f,\nu)$ is distributed as a non-central t distribution with $f$ d.f. and non-central parameter $\nu$, and $t_{f,p}$ is the $p$-th percentile of $t_f$.  
To simplify the calculation, we approximate the power by  
\begin{eqnarray}\label{power}
\begin{aligned}
P  = \text{Pr}(\hat\tau_p^*\geq  t_{f,1-\frac{\alpha}{2}} -\sqrt{n}\lambda)+\text{Pr}(\hat\tau_p^*\leq  -t_{f,1-\frac{\alpha}{2}} -\sqrt{n}\lambda) \approx  \text{Pr}(\hat\tau_p^*\leq \sqrt{n}|\lambda| - t_{f,1-\frac{\alpha}{2}} ),
\end{aligned}
\end{eqnarray}
where $\hat\tau_p^*=(\hat{\tau}_p -\tau_p)/\sqrt{\widehat{\textrm{var}}_{\text{kr}}(\hat\tau_p)}$  approximately follows a t distribution with $f$ d.f.,
and  $$\sqrt{n}\lambda = \frac{\tau_p}{\sqrt{\widehat{\textrm{var}}_{\text{kr}}(\hat\tau_p)}} \approx \frac{\tau_p}{\sqrt{n^{-1}\varpi_\tau(n)}}.$$
  For a scalar fixed effect, Kenward and Roger \cite{1997:kenward} used the Satterthwaite-type d.f., which is a random variable (could be larger than $n_1-q$), and  difficult to evaluate at the design stage.
We approximate $f$ by $$f(n)=(n_1-q)f_o= (n \bar\pi_1 - q) f_o,$$ 
where $$f_o= \frac{(\sum_{j=1}^p l_{pj}^2\sigma_j^2)  \varpi_{x_1} }{\sum_{j=1}^p l_{pj}^2\sigma_j^2  \varpi_{x_j} } \approx  
 \frac{( \sum_{j=1}^p l_{pj}^2\sigma_j^2)  \varpi_{\pi_1} }{\sum_{j=1}^p l_{pj}^2\sigma_j^2  \varpi_{\pi_j} } $$ can be roughly interpreted as the fraction 
of observed information among subjects retained at visit $1$.

There is no closed-form solution for the sample size needed to achieve a desired power $P$.
We propose a two-step procedure to approximate the required size.
Inverting equation \eqref{power} yields  
\begin{equation}\label{size1}
n_{\text{\tiny tar}} = (t_{f,1-\frac{\alpha}{2}}+t_{f,\text{P}})^2\frac{ \varpi_\tau(n)}{\delta_p^2}=\frac{(t_{f,1-\frac{\alpha}{2}}+t_{f,\text{P}})^2}{\lambda^2}.
\end{equation}
In large samples, we can approximate $t_{f,p}$ by the  $p$th percentile $z_p$ of  $N(0,1)$, and approximate
 $\varpi_\tau(n)$  by $\varpi_{\tau}^*[1+q^* (\sum_{j=1}^p b_j/\bar\pi_j)/n]$, 
where $\varpi_{\tau}^* = \sum_{j=1}^p l_{pj}^2\sigma_j^2  \varpi_{\pi_j}$ and $b_j=l_{pj}^2\sigma_j^2\varpi_{\pi_j}/\varpi_{\tau}^*$.
In step $1$, we find the sample size based on the normal approximation
\begin{equation}\label{lowbound}
n_l=(z_{1-\frac{\alpha}{2}}+z_{\text{P}})^2  \frac{\varpi_{\tau}^*}{\delta_p^2} +q^*\sum_{j=1}^p\frac{b_j}{\bar\pi_j} \approx (z_{1-\frac{\alpha}{2}}+z_{\text{P}})^2  \frac{\varpi_{\tau}^*}{\delta_p^2} +\frac{q^*}{\bar\pi_p},  
\end{equation}
 and it generally provides a sharp lower bound on the required size. If the retention rate is low (e.g. $\bar{\pi}_p<0.5$), we use the first equality in equation \eqref{lowbound}.
In the implementation, we round $n_l$ up to the nearest integer.
 In step $2$, we use the approximation
\begin{equation}\label{upbound}
 n_u=(t_{f(n_l),1-\frac{\alpha}{2}}+t_{f(n_l),\text{P}})^2 \frac{\varpi_\tau(n_l)}{\delta_p^2},
\end{equation}
and it usually  provides a very good  estimate of the required size. We may approximate $n_u$ by 
\begin{equation}\label{upbound2}
 n_u^*\approx (t_{f(n_l),1-\frac{\alpha}{2}}+t_{f(n_l),\text{P}})^2 \frac{ \varpi_\tau^*}{\delta_p^2} +q^*\sum_{j=1}^p\frac{b_j}{\bar\pi_j}
\approx (t_{f(n_l),1-\frac{\alpha}{2}}+t_{f(n_l),\text{P}})^2 \frac{ \varpi_\tau^*}{\delta_p^2} +\frac{q^*}{\bar{\pi}_p}.
\end{equation}
We recommend using equation \eqref{upbound} or equation \eqref{upbound2} to estimate the sample size, and using equation \eqref{power0} to evaluate the nominal power at a given sample size
since equation \eqref{power0} is more accurate than equation \eqref{power} when $f$ is extremely small (e.g. $f< 12$).
If the nominal power at $n_u$ or $n_u^*$ is not close to the targeted power, one may increase or decrease the sample size 
to achieve the desired power.

We also try a slight variation of the proposed size estimation procedure by increasing $n_l$ defined in equation \eqref{lowbound} by $2$, but all other steps remain the same. More details are provided 
in simulation $1$ reported in Section 3.2.1 below.

\begin{table}
 \centering
 \def\~{\hphantom{0}}
 \begin{minipage}{150mm}
\caption{Calculated sample sizes and  simulated  power based on $10,000$  replications for testing $H_0:\tau_4=0$ in simulation $1$: 
$^a$ The sample size is evaluated using equation \eqref{upbound}. The size estimates from equations \eqref{lowbound} and  \eqref{upbound2} are  presented for the purpose of comparison;
$^b$ The difference in sample size between two arms is $0$ for even $n$ and $1$ for odd $n$;
$^c$ The nominal power is evaluated  as $\text{Pr}( t(f(n),\sqrt{n}|\tau_4|/\sqrt{ \varpi_\tau(n_l)}) \geq  t_{f(n),0.975})$ based on equation \eqref{power0}.
 The difference in nominal power estimated by equation \eqref{power0} and equation \eqref{power} is $<0.2\%$ in all cases.
}\label{powersize}
\begin{tabular}{lcccccccccc} \\\hline 
 true covariance          & &  \multicolumn{3}{c}{estimated sample size$^a$} & total &  \multicolumn{2}{c}{power ($\%$)} \\ \cline{3-5}\cline{7-8}    
structure & $\tau_4$ & $n_l$ & $n_u^*$ & $n_u$  & size $n$ $^b$ & nominal$^c$ & simulated \\ \hline
1: Unstructured    &  $-12$ & $17$ & $19.4$ & $20.4$ & $21$ & $91.86$ & $91.48$\\
    &  $-8$ & $36$ & $38.1$ & $38.6$ & $39$ & $90.49$ & $89.99$  \\
     &  $-4$ & $139$ & $140.8$ & $141.0$ & $142$ & $90.22$ & $90.47$\\
2: CS   &  $-12$ & $19$ & $21.1$ & $22.7$  & $23$ & $91.36$ & $90.84$  \\
   &  $-8$   &   $40$ & $41.8$  & $42.7$  & $43$ & $90.36$ & $89.74$   \\
 &   $-4$ & $153$ & $154.8$ & $155.5$ & $156$ & $90.11$ & $90.22$ \\
3: AR(1) & $-12$ & $17$ & $19.6$ & $20.7$ & $21$ & $91.53$ & $91.10$ \\
 & $-8$   &  $36$ & $38.4$ & $39.0$  & $39$ &  $90.24$ & $89.75$ \\
 & $-4$ & $140$ & $141.8$ & $142.1$ & $143$ & $90.21$ & $90.87$ \\
4: Toeplitz & $-12$ & $15$   & $17.6$ & $18.7$ & $19$ & $91.81$ & $91.86$\\                                                                                         
 & $-8$ &  $32$ & $33.9$ & $34.3$ & $35$ & $90.81$ & $90.31$ \\
 & $-4$ &  $122$ & $123.7$ & $123.9$ & $124$ & $90.04$ & $89.71$ \\\hline
\end{tabular} 
\end{minipage}
\vspace*{6pt}
\end{table}

\subsection{Numerical examples}
\subsubsection{Simulation 1.}
In an antidepressant trial analyzed in Section \ref{exam1}, the active treatment is  significantly better than placebo in reducing depression.
It might be of interest to design a  new study  to assess the effect of a similar  compound on $\text{HAMD}_{17}$. 
Suppose $p=4$, $y_{i0}\stackrel{i.i.d}{\sim} N(17.9,5.5^2)$, $\mu_{i1}=3.3+0.72y_0 +0.1g_i$, $\mu_{i2}=2.7+0.69y_0-1.5g_i$,
$\mu_{i3}=2.9+0.61y_0-2.3g_i$, $\mu_{i4}=1+0.67y_0+\tau_4 g_i$,
$$\begin{bmatrix} y_{i1} \\
 y_{i2}\\
y_{i3}\\
y_{i4}\end{bmatrix} \sim N_4\left( \begin{bmatrix} \mu_{i1} \\ \mu_{i2} \\ \mu_{i3} \\ \mu_{i4} \\\end{bmatrix}, 
 \begin{bmatrix} 19.68  & 16.45  &  15.39  &   16.36 \\
             16.45    &    34 &     25.34 &    26.13 \\
             15.39    & 25.34   &  38.44 &    33.91\\
            16.36    &  26.13 &     33.91  &   45.28  \\ \end{bmatrix}\right),$$
and the retention rate is $(\pi_{01},\ldots,\pi_{04})= (1,0.92,0.86,0.74)$ and $(\pi_{11},\ldots,\pi_{14})=(1,0.93,0.87,0.76)$.
These parameters are specified based roughly on the MMRM analysis of the antidepressant trial. Note that the  sample size
depends on the retention rates, the treatment effect $\tau_4$ at the last visit, and $\Sigma$. Other parameters are specified in order to simulate the data.
For the purpose of illustration, we set
$\tau_4=-4$, $-8$ or $-12$, and three alternative covariance structures are also considered:
1)   a compound symmetry (CS) structure: $\Sigma_{kk}=45$ and $\Sigma_{kj}=15$ if $k\neq j$; 2) 
 an autoregressive (AR(1)) structure $\Sigma_{jk}= 45\times 0.8^{|j-k|}$; 3)  a Toeplitz  structure $\Sigma_{jk}=40 -6|j-k|$.
For CS and AR(1) structures, the analytic expressions for the LDL decomposition provided in the appendix can be used in the calculation.

We calculate the sample  size necessary to achieve $90\%$ power at  a two-sided significance level of $\alpha=0.05$ using equation \eqref{upbound}.
In each case, $10000$ datasets are simulated and  analyzed using model \eqref{mixed1} and  KR variance estimate. 
There is about $95\%$ chance that the simulated power lies within $0.6\%$ (standard error $\approx \sqrt{0.9*0.1/10000}=0.3\%$) of the true power.
The result is summarized in Table \ref{powersize}. 
The difference between the simulated  and nominal power is $<1\%$ in all cases.

Empirical evidence indicates that the normal approximation based on equation \eqref{lowbound} usually underestimates the sample size by at least $2$. 
We try a slight variation of the sample size estimation procedure by increasing $n_l$ by $2$ (all other steps remain the same).
This modification makes a difference mainly in small samples (i.e. $\tau_4=-12$) in that the  sample size estimate is reduced by $1$. We still recommend 
the procedure described in Section $3.1$ since it is generally desirable to use a conservative size estimate.

\subsubsection{Simulation 2.}
This simulation illustrates that the power and sample size depend on the number of covariates.  
The covariates  consist of the baseline outcome $y_{i0}$, treatment status $g_i$ and a categorical prognostic factor $A$ with three levels.  
Suppose  the status of the prognostic factor is $A_i$ for subject $i$,
$y_{i0}\stackrel{i.i.d}{\sim} N(\eta_{_{A_i}}+17.9,5.5^2)$, and
$$\begin{bmatrix} y_{i1} \\
 y_{i2}\\
y_{i3}\\
y_{i4}\end{bmatrix} \sim N_4\left( \begin{bmatrix} \eta_{_{A_i}}+\mu_{i1} \\ \eta_{_{A_i}}+ \mu_{i2} \\ \eta_{_{A_i}}+\mu_{i3} \\ \eta_{_{A_i}}+\mu_{i4} \\\end{bmatrix}, 
\Sigma\right),$$
where $\eta_1=0$, $\eta_2=-0.5$, $\eta_3=0.5$, and $\mu_{ij}$'s are defined in simulation $1$.
 We assume that the effect of the prognostic factor  is constant over time, and that
each subject is in level $1$, $2$ and $3$ of the prognostic factor  with probability $0.3$, $0.4$, and $0.3$ respectively.
Other assumptions  are the same as that in simulation $1$.  In the MMRM analysis, the fixed effects include  visit, 
$y_{i0}\times \text{visit}$, $A\times \text{visit}$ and $\text{treatment}\times \text{visit}$ interactions.

The result is displayed in Table \ref{powexam3}. The required size in simulation $2$ is larger than that in simulation  $1$ under the same assumption on the treatment effect and 
residual covariance matrix $\Sigma$ due to the increase in the number of baseline covariates.
The difference between the simulated  and nominal power is $<1\%$ in all cases. 
Overall,  the sample size is underestimated by the normal approximation approach given in equation \eqref{lowbound} especially at $\tau_4=-12$, and
the proposed two-step sample size calculation method performs well when the d.f. in equation \eqref{upbound} is larger than $f(n_l)\geq 12$.

\begin{table}
 \centering
 \def\~{\hphantom{0}}
 \begin{minipage}{150mm}
\caption{Calculated sample sizes and  simulated  power based on $10,000$  replications  for testing $H_0:\tau_4=0$  in simulation $2$: 
$^a$ The sample size is evaluated using equation \eqref{upbound}. The size estimates from equations \eqref{lowbound} and  \eqref{upbound2} are  presented for the purpose of comparison;
 $^b$ The difference in sample size between two arms is $0$ for even $n$ and $1$ for odd $n$;
$^c$ The nominal power is evaluated as $\text{Pr}( t(f(n),\sqrt{n}|\tau_4|/\sqrt{ \varpi_\tau(n_l)}) \geq  t_{f(n),0.975})$ based on equation \eqref{power0}.
}\label{powexam3}
\begin{tabular}{lcccccccccc} \\\hline 
 true covariance          & &  \multicolumn{3}{c}{estimated sample size $^a$} & total &  \multicolumn{2}{c}{power ($\%$)} \\ \cline{3-5}\cline{7-8}    
structure & $\tau_4$ & $n_l$ & $n_u^*$ & $n_u$  & size $n$ $^b$ & nominal$^c$ & simulated \\ \hline
1: Unstructured    &  $-12$ & $20$ & $21.9$ & $23.5$ & $24$ & $91.51$ & $91.67$\\
                          &  $-8$ & $39$ & $40.7$ & $41.2$ & $42$ & $90.70$ & $90.91$  \\
                          &  $-4$ & $142$ & $143.4$ & $143.4$ & $144$ & $90.13$ & $89.65$\\
2: CS   &  $-12$ & $21$ & $23.8$ & $27.1$  & $28$ & $92.30$ & $92.97$  \\
          &  $-8$   &   $42$ & $44.5$  & $45.9$  & $46$ & $90.27$ & $90.01$   \\
            &   $-4$ & $155$ & $157.5$ & $158.2$ & $159$ & $90.17$ & $89.99$ \\
3: AR(1) & $-12$ & $20$ & $22.0$ & $23.9$ & $24$ & $91.09$ & $91.12$  \\
             & $-8$   &  $39$ & $41.0$ & $41.7$  & $42$ &  $90.42$ & $90.67$ \\
             & $-4$ & $143$ & $144.4$ & $144.5$ & $145$ & $90.12$ & $89.74$ \\
4: Toeplitz & $-12$ & $18$   & $20.0$ & $21.9$ & $22$ & $91.31$ & $91.52$\\                                                                                         
               & $-8$ &  $35$ & $36.4$ & $36.9$ & $37$ & $90.24$ & $89.96$ \\
               & $-4$ &  $125$ & $126.4$ & $126.3$ & $127$ & $90.17$ & $90.52$ \\\hline
\end{tabular} 
\end{minipage}
\vspace*{6pt}
\end{table}

\subsubsection{Simulation 3.} So far, we assume the data are normally distributed. The third simulation is conducted to assess the performance of the proposed method for non-normal data in small to moderate samples
($\tau_4=-8$ or $-12$).
The data are simulated from the multivariate skew-normal distribution \cite{azzalini:1996} and multivariate t distribution. 
For the purpose of comparison, we set the covariance matrix of  $(y_{i1},\ldots,y_{i4})$ and 
the treatment effect at each visit to be identical to that in simulation $1$. The nominal power and sample size estimates will be same as that in simulation $1$. 

In the multivariate t distribution, we generate the data as $y_{ij}=\mu_{ij}+\epsilon_{ij}/\sqrt{u_i/d}$, where $u_i \stackrel{i.i.d.}{\sim} \chi_d^2$ follows a chi-square distribution with $d$ d.f., and
$(\epsilon_{i1},\ldots,\epsilon_{i4})'$ is normally distributed with zero mean vector and covariance matrix $\frac{d-2}{d}\Sigma$. 
The covariance matrix of $(y_{i1},\ldots,y_{i4})$ is $\Sigma$. We set $d=6$ or $10$. The skewness of $y_{ij}$ is 0. The (excess) kurtosis of $y_{ij}$ is $6/(d-4)=3$ at $d=6$ and $1$ at $d=10$. 

The skew-normal distribution \cite{azzalini:1996} can be used to model data that are mildly to moderately skewed. 
Let $R=D^{-1}\Sigma D^{-1}$ be the correlation matrix corresponding to $\Sigma$, and $R_2= (1-\kappa^2)^{-1}[a R- b \onev_p\onev_p']$,
where $\kappa$ is a prespecified scalar value in the interval $(-1,1)$, $a=1-2 \kappa^2/\pi$, $b=(1-2/\pi)\kappa^2$, and $D=\text{diag}(\Sigma_{11},\ldots,\Sigma_{pp})$. 
The data are generated as  
$y_{ij}=\mu_{ij}+  \sqrt{\Sigma_{jj}/a} \,\epsilon_{ij}$, where $\epsilon_{ij}= \kappa e_i + \sqrt{1-\kappa^2} \,\tilde\epsilon_{ij}$,
$e_i \stackrel{i.i.d.}{\sim} N(0,1)$, and $(\tilde\epsilon_{i1},\ldots,\tilde\epsilon_{ip})'\sim N_p(\zerov, R_2)$.  
The covariance matrix of $(y_{i1},\ldots,y_{i4})'$ is $\Sigma$. The marginal distribution of $\tilde\epsilon_{ij}$ is $N(0,1)$, and
the skewness of $\epsilon_{ij}$ ($y_{ij}$ has the same skewness as $\epsilon_{ij}$) ranges from $-0.995$ to $0.995$ as $\kappa$ changes from $-1$ to $1$.
We set $\kappa=0.8$ and $0.9$. No result is produced for the CS covariance  structure at $\kappa=0.9$ since the corresponding  $R_2$ is not positive definite. 

The result is presented in Table \ref{powersizeskew}. 
In general, the type I error is close to the nominal $5\%$ level except in few cases  under the t distribution with $6$ d.f., where
the t test with KR adjusted variance provides slightly conservative control of the type I error. 
The performance on the power estimate is almost as good as that under the normality assumption in simulation $1$. 
The simulated power is within $1.2\%$ of the nominal power in all cases. 
Additional simulation (results not shown) is conducted under the skew-normal distribution at  $\kappa=-0.8, -0.9$ and $\pm0.95$, and different covariance matrices 
may be used to ensure that $R_2$ is positive definite.
All simulations indicate that the proposed power and sample size estimation procedure is fairly  robust to mild or moderate deviations from non-normality even when $n=20$.

\begin{table}
 \centering
 \def\~{\hphantom{0}}
 \begin{minipage}{150mm}
\caption{Calculated sample sizes and  simulated  power based on $10,000$  replications for testing $H_0:\tau_4=0$ under multivariate t and skew-normal distributions  in simulation $3$: $^a$
The sample size and nominal power estimates are identical to that in simulation $1$;
 $^b$ Type I error (Type1) is evaluated at $\tau_4=0$; $^c$ Simulated power (SIM) is evaluated at $\tau_4=-8$ and $-12$.
}\label{powersizeskew}
\begin{tabular}{lrrccc@{\extracolsep{3pt}}c@{}c@{\extracolsep{3pt}}c@{}c@{\extracolsep{3pt}}c@{}cccccc} \\\hline 
                                 & & total & nominal & \multicolumn{4}{c}{skew-normal distribution ($\%$)} & \multicolumn{4}{c}{multivariate t distribution ($\%$)}\\\cline{5-8}\cline{9-12}
 true covariance   & & size$^a$ & power $^a$& \multicolumn{2}{c}{$\kappa=0.8$}  &  \multicolumn{2}{c}{$\kappa=0.9$}  & \multicolumn{2}{c}{$10$ d.f.} & \multicolumn{2}{c}{$6$ d.f.} \\\cline{5-6}\cline{7-8}\cline{9-10}\cline{11-12}
structure & $\tau_4$ &   $n$  & ($\%$) & Type1$^b$ & SIM$^c$ & Type1 & SIM & Type1 & SIM & Type1 & SIM \\ \hline
1: Unstructured    &  $-12$ &  $21$& $91.86$ & $4.97$  & $91.26$ &  $4.84$  & $91.02$  & $5.21$ & $90.80$& $4.39$ & $90.87$ \\
    &  $-8$ & $39$  & $90.49$  & $5.05$& $90.07$ & $5.10$& $90.32$  & $4.97$ & $89.52$& $4.83$ & $90.25$ \\
2: CS   &  $-12$ &  $23$ & $91.36$ & $4.86$  & $90.56$  & $-$  & $-$ & $5.09$ & $90.30$& $4.72$ & $90.18$\\
   &  $-8$   &    $43$ & $90.36$ & $4.84$  & $89.66$ & $-$  & $-$  & $5.28$ & $89.63$ & $5.06$ & $89.82$ \\
3: AR(1) & $-12$ &  $21$ & $91.53$ & $4.89$  & $90.91$ & $4.99$  & $90.72$ & $5.13$ & $90.54$& $4.35$ & $90.45$\\
 & $-8$   &  $39$ & $90.24$ &  $5.06$  & $89.92$ &  $4.93$  & $89.91$ & $4.92$ & $89.28$& $4.86$ & $90.09$\\
4: Toeplitz & $-12$ &  $19$ & $91.81$ & $5.24$  & $90.80$ & $4.93$  & $90.88$& $4.91$ & $91.07$& $5.11$ & $91.13$\\                                                                                         
 & $-8$ &   $35$ & $90.81$ & $5.01$  & $90.26$ & $4.82$  & $90.84$ & $5.30$ & $89.69$& $4.99$ & $90.36$\\\hline
\end{tabular} 
\end{minipage}
\vspace*{6pt}
\end{table}

\section{Discussion}
\setlength{\parindent}{3ex}
We derive the closed-form REML estimator and compare several variance estimators for fixed effects  in  MMRM when the missing data pattern is monotone. 
The bias in the ML and REML parameter estimators is assessed. For monotone missing data, the MMRM yields the same treatment effect as the cLDA \cite{lu:2010},
but it makes less assumption on the distribution of baseline outcomes.
One referee raises concern over the terminology {\it MMRM}  since there is no random effect
in model \eqref{mixed1}. We partially agree with the referee, 
but this terminology  is commonly used in the pharmaceutical industry \cite{aiddiqui:2009, aiddiqui:2011, lu:2008}. 

A two-step sample size determination procedure is proposed  for the KR adjusted t test for comparing two treatments in MMRM. 
Simulation demonstrates its good performance even in small samples. 
When the d.f. given in equation  \eqref{upbound} is  $f(n_l) < 12$, the power formula \eqref{power0} still works reasonably well, but the 
required size can be overestimated. The main reason is because  the normal approximation approach
generally underestimates the required  size and hence the d.f., leading to non-negligible inflation in  $t_{f,1-\alpha/2}+t_{f,\text{P}}$ and $\varpi_\tau(n)$ 
in small samples. When $f(n_l) < 12$, we would recommend using a numerical (e.g. bisection) method
to find the size based on equation \eqref{power0}, which is the smallest integer at which the power  reaches the pre-specified level. 
Simulation indicates that the proposed method  is quite robust to mild or moderate deviations from non-normality.
However, this robustness property may not hold under severe non-normality even for the one sample t test \cite{boos:2000}.
It is always prudent to run a simulation study to verify the power and sample size estimate especially when the data are highly non-normal or when the sample size is small.
We assume balanced baselines  across groups. 
If the baselines are imbalanced, we may replace equation \eqref{expvar1} by  
$$\text{E}(\Delta_{t}' S_{x_{t}}^{-1}\Delta_{t})\approx \frac{\varpi_{\delta_t}(n)\varpi_{\pi_t} }{n}+ \frac{D}{n\bar\pi_t-q-1}
\text{ and  } \text{E}(V_{x_t})=\frac{\varpi_{x_t}(n)}{n}+ \frac{D}{n\bar\pi_t-q-1},$$
where $\muv_d$ is the difference in mean of $\xv_{bi}$ between two arms, $\Sigma_x$ is the covariance matrix of $\xv_{bi}$ and 
$D = \muv_d'\Sigma_x^{-1}\muv_d$.

In a companion paper, we investigate the power of treatment comparison in an unstratified randomized trial,
and  the power for testing treatment effect and treatment by stratum interaction in a stratified  trial using analysis of covariance (ANCOVA). 
For the MMRM analysis of a stratified trial  with $h$ strata, we may replace equation \eqref{expvar1} by  
\begin{equation}\label{expvars}
\varpi_{\delta_t}(n)= \frac{q^*}{n\bar\pi_t-q-1}    \text{ and }
   \text{E}(V_{x_t})\approx \frac{\varpi_{x_t}(n)}{n}= \frac{ \varpi_{\pi_t}[1+\varpi_{\delta_t}(n)] }{n},
\end{equation} 
where  $q^*$ is the dimension of covariates excluding intercept, treatment status, and pre-stratification factors,
 $(x_{i_1},\ldots,x_{i_{h-1}})$ are the indicator variables for strata,  $\xv_{i}= (1, x_{i_1},\ldots,x_{i_{h-1}}, \xv_{bi}',g_i)'$ and $q=q^*+h+1$.
Equation \eqref{expvars} is derived by assuming constant treatment  effect, treatment  allocation ratio and retention rate across strata.

Simulation $2$ indicates that  including the correction term $\varpi_{\delta_t}$ in $\varpi_{x_t}$ in equations \eqref{power0} and \eqref{power}  and the last term in equations \eqref{lowbound} and  \eqref{upbound}
can greatly improve the accuracy of the power  and sample size estimate when $q^*$ is  large or $n$ is relatively small.
Covariate selection is critical in small trials. In the power calculation, $\Sigma=L\Lambda L'$ is the variance of the outcome unexplained by the covariates.
 Inclusion of important covariates can reduce the residual variance $\Sigma$, and increase the power of the analysis.
However, if the covariates  are unrelated or only weakly related to the outcome, both the precision of the treatment effect estimate (by equation \eqref{expvar1})
and the d.f. in the t test may decrease, resulting in reduced power.

 It can be challenging to specify $\Sigma$  at the design stage of a trial. 
One strategy is to assume a structured covariance matrix that represents the best guess about the true covariance matrix  to reduce the number of nuisance parameters
 in the  sample size calculation \cite{lu:2009}, but the data are still analyzed by model \eqref{mixed1}. Furthermore,  blinded size reassessment procedure may be used to
 re-estimate the variance parameters and adjust the  size  based on interim
blinded data, in which the treatment effect is generally small and thus ignored in estimating the variance parameters \cite{tang:2015c}. 

We  assume monotone missing pattern in the power and sample size calculation. In practice, the trials generally contain only a small amount of intermittent missing data. For example,
in the anti-depressant trial analyzed in Section $2.4$, only one out of $172$ subjects has intermittent missing values. In the presence of non-monotone missing data,
sample size can be calculated in a slightly conservative way by excluding subjects with intermittent missing data or by excluding data collected after the first missing visit.
One needs to adjust the retention rates before applying  the power and sample size formulae derived in section $3.1$. If  a large amount of intermittent missing data is expected,
one may use simulation to find the appropriate sample size. The above approach provides an upper size bound, and
a lower size bound can be obtained by pretending there are no intermittent missing data.

One limitation of the proposed method is that the data are assumed to be missing at random in the MMRM. Sensitivity analyses under the nonignorable missingness have become increasingly popular in 
new drug applications \cite{kenward:2015, 2016:tangb, 2016:tang}. Sample size formulae for such sensitivity analysis may be obtained by using the analytical expressions for the MI treatment effect estimate and 
the associated variance derived in \cite{2016:tangc}.

We have focused on the longitudinal continuous outcome. For other types of longitudinal outcomes (e.g. binary or ordinal), one may use the methods described in \cite{liu:1997, rochon:1998}
to compute the sample size for the tests based on the generalized estimating equations (GEE).  However, these methods may not be suitable for small samples \cite{lu:2007}.

The analytic REML solution is useful in statistical computing. For example, we find the default REML estimation
algorithm used by SAS Proc mixed fails to converge in about $5\%$ monotone datasets at $n=15$ in a simulation, but the REML estimate
can be easily obtained using Lemma $1$. For near-monotone datasets,  the convergence of the REML algorithm may be  accelerated
by using the REML estimate excluding subjects with intermittent missing data as the initial parameter values.
For non-monotone data, Liu \cite{liu:1999} proposed a MEM algorithm to find the ML estimate. 
It is also possible to develop a MEM algorithm for the REML estimation, and the details will be presented elsewhere.

\begin{center}
{\bf Acknowledgements}
\end{center}
We would like to thank  the associate editor and  two  referees for their helpful suggestions  that improve the quality of the manuscript. 
 
\appendix
\section{Appendix: technical details}\label{proof}

 \begin{proof}{of Lemma \ref{solreml}:}
(a) 
Let  $U_k$, $L_k=U_k^{-1}$, $\Lambda_k$ and $\Sigma_k = L_k\Lambda_k L_k'$ denote respectively the leading $k\times k$ submatrix 
 of  $U$, $L=U^{-1}$, $\Lambda$ and $\Sigma=L\Lambda L'$. Then
$W_i=\tilde{I}_{r_i}\otimes \xv_i$,   
$W_i'\Sigma_{r_i}^{-1}W_i= (\sum_{s=1}^{r_i} \tilde\betav_s'\tilde\betav_s/\sigma_s^2) \otimes (\xv_i'\xv_i)$,
 $\sum_i W_i'\Sigma_{r_i}^{-1}W_i= \sum_{s=1}^p [(\tilde\betav_s'\tilde\betav_s) \otimes (X_{o_s}'X_{o_s}/\sigma_s^2)]$,
 $(L'\otimes I_p) (\sum_i W_i'\Sigma_{r_i}^{-1}W_i) (L\otimes I_p) =V_w^{-1}$, where 
$\tilde{I}_k$ is formed by the first $k$ row of $I_p$, and
$\tilde\betav_s=(-\beta_{s1},\ldots,-\beta_{ss-1},1,0,\ldots,0)'$.  Thus (a) holds.\\
(b) We can prove (b) by showing  $\sum_i\yv_i \Sigma_{r_i}^{-1}\yv_i = \sum_{j=1}^p (\sigma_j^{-2}\sum_{k=1}^{n_j} \underline{y}_j^2)$,
 $\sum_i \log{|\Sigma_{r_i}|} -\log{|\sum_i W_i'\Sigma_{r_i}^{-1}W_i|}= \sum_{j=1}^p (n_j-q)   \log(\sigma_j^2)$,                                                         
$\sum_{i=1}^n W_{i}'\Sigma_{r_i}^{-1} \yv_{io}  = (U'\otimes I_p) \begin{bmatrix} 
               \sigma_1^{-2} \sum_{i=1}^{n_1} \xv_i'\underline{y}_{i1} \\
               \ldots\\                
      \sigma_p^{-2} \sum_{i=1}^{n_p} \xv_i'\underline{y}_{ip}
      \end{bmatrix}$,  and the last term of equation \eqref{reml1} equals  $-\sum_{j=1}^p[ \sigma_j^{-2} (\sum_{i=1}^{n_j} \xv_i'\underline{y}_{ij})' (X_{o_j}'X_{o_j})^{-1}(\sum_{i=1}^{n_j} \xv_i'\underline{y}_{ij})]$. \\
(c) Maximizing equation \eqref{reml2} yields  $(\hat\betav_j,\hat\sigma_{j,re}^2)$'s, and
 $\underline\alphav_j$'s are evaluated  at the  REML estimator $(\hat\betav_j,\hat\sigma_{j,re}^2)$'s. 
We can prove
$\hat\thetav_{j,re}=\hat\thetav_j$ by using 
\begin{equation}\label{matrixinv}
\begin{bmatrix} x'x & x'z \\
                           z'x & z'z 
   \end{bmatrix}^{-1} = \begin{bmatrix}  (x'x)^{-1} +\eta V_{z} \eta' & -\eta V_{z} \\
                                  -V_{z}\eta' & V_{z} \\
   \end{bmatrix},
\end{equation}
where $V_{z}=[z'(I-x(x'x)^{-1}x')z]^{-1}$, and $\eta=(x'x)^{-1}x'z$.\\
(d) Let $\underline{\vec{Y}}_{o_j}= (\underline{Y}_{o_{j1}},\ldots, \underline{Y}_{o_{j,j-1}})= \vec{Y}_{o_j} U_{j-1}'$. 
By equation \eqref{factor0}, $\underline{Y}_{o_{jk}} \sim N(X_{o_j}\underline{\alphav}_k, \sigma_k^2 I_{n_j})$, and 
$\underline{Y}_{o_{jk}}$'s ($k=1,\ldots,j-1$) are independent. Lemma \ref{solreml}(e) holds since
$\text{E}(\underline{\vec{Y}}_{o_j}'Q_j\underline{\vec{Y}}_{o_j})=(n_j-q)\Lambda_{j-1}$ and
 $\text{E}(\vec{Y}_{o_j}'Q_j\vec{Y}_{o_j})=(n_j-q)\Sigma_{j-1}$.
\end{proof}

\begin{proof}{\bf of equation \eqref{varalphadiff}:}
A little algebra shows that $\hat\alphav_k= \sum_{t=1}^{k-1}\hat\beta_{kt} \hat{\alphav}_t +\hat{\underline\alphav}_k$, $\tilde\alphav_k= \sum_{t=1}^{k-1}\beta_{kt} \tilde{\alphav}_t +\tilde{\underline\alphav}_k$, 
$\hat\alphav_k- \tilde\alphav_k = d_k+e_k + \sum_{t=1}^{k-1} \beta_{kt} (\hat\alphav_t - \tilde\alphav_t)$, and 
\begin{equation}\label{varde}
 (\hat\alphav_1 - \tilde\alphav_1,\ldots,\hat\alphav_k- \tilde\alphav_k)'= L_{k} (d_1+e_1,\ldots,d_k+e_k)',
\end{equation}
where $\hat{\underline\alphav}_{kt}^*=(X_{o_k}'X_{o_k})^{-1}X_{o_k}'{\underline{Y}}_{o_{kt}}$, $\zeta_k=L_{k-1}'(\hat\betav_k-\betav_k)$,
$\tilde{\underline\alphav}_k-\hat{\underline\alphav}_k=(X_{o_k}'X_{o_k})^{-1}X_{o_k}'\vec{Y}_{o_k}(\hat\betav_k-\betav_k)= (\hat{\underline\alphav}_{k1}^*,\ldots, \hat{\underline\alphav}_{kk-1}^*) \zeta_k$,
$d_1=e_1=\zerov$,
 $d_k= \sum_{t=1}^{k-1}(\hat\beta_{kt}-\beta_{kt})\tilde{\alphav}_t + (\hat{\underline\alphav}_k-\tilde{\underline\alphav}_k)= (\tilde{\underline{\alphav}}_1-\hat{\underline\alphav}_{k1}^*,\ldots,\tilde{\underline\alphav}_{k-1}-\hat{\underline\alphav}_{kk-1}^*)\zeta_k$ and
$e_k= \sum_{t=1}^{k-1}(\hat\beta_{kt}-\beta_{kt})(\hat{\alphav}_t-\tilde{\alphav}_t)=(d_1+e_1,\ldots,d_{k-1}+e_{k-1})\zeta_k$ at $k\geq 2$. Note that 
$d_k$'s, $e_k$'s and $\tilde{\underline\alphav}_k$'s are independent, $\text{var}(d_k) = \sum_{t=1}^{k-1}  \sigma_t^2\omega_{kt} D_{kt} $, and $\text{E}(d_k)=\text{E}(e_k)=\zerov$.
Thus $\text{cov}(\tilde{\alphav}, \hat\alphav-\tilde\alphav)=\zerov$. Let $\epsilonv_j=\hat\alphav_j-\tilde\alphav_j$ and $\epsilonv= (\epsilonv_1',\ldots,\epsilonv_p')'$. 
By equation \eqref{varde}, we have $\text{var}(\epsilonv)= (L\otimes I_q) V_{de} (L\otimes I_q)'$ and $\text{var}(\tilde\alphav_p -\hat\alphav_p)=\sum_{j=2}^p l_{pj}^2 (V_{dj}+V_{ej})$,
 where $V_{de}=\text{diag}(\zerov,V_{d2}+V_{e2},\ldots, V_{dp}+V_{ep})$.
\end{proof}

\begin{proof}{\bf of equation \eqref{biasasy}:} Let $g_j=\sigma_j^2 l_{pj}^2$ and $c_j=\text{E}(\hat{g}_j)-g_j$. Then 
$\text{E}(\hat\Phi_p)-\Phi_p=  \sum_j c_j  (X_{o_j}'X_{o_j})^{-1}$.
We  evaluate $c_j$ using the second-order Taylor series  approximation
$$ c_j \approx \frac{\partial g_j}{\partial \sigma_j^2} \text{E}(\hat\sigma_j^2-\sigma_j^2) +\frac{1}{2} \frac{\partial^2 g_j}{\partial l_j\partial l_j} \text{E}(\hat{l}_j-l_j)^2
 = \sigma_j^2 \text{var}(l_{pj})  - \frac{j-1}{n_j-q}\sigma_j^2 l_{pj}^2, $$
by noting $\text{E}(\hat{l}_j-l_j)=0$, $\text{E}(\hat{l}_j-l_j)(\hat\sigma_j^2-\sigma_j^2)=0$ and $\partial ^2 g_j/\partial \sigma_j^2 \partial\sigma_j^2=0$.
Since $\partial(l_{pj})/\partial\betav_k = l_{pk} (l_{1j},\ldots,l_{k-1,j})'$ [it is $\zerov$ if $j\geq k$] and $(l_{1j},\ldots,l_{k-1,j})\text{var}(\betav_k)(l_{1j},\ldots,l_{k-1,j})'=\omega_{kj}$,
 we have $\text{var}(l_{pj})= \sum_{k=j+1}^p l_{pk}^2 \omega_{kj}$ and equation \eqref{biasasy} holds.
\end{proof}

\flushleft{\bf Analytic expressions for  LDL decomposition of CS and AR(1) covariance}\\
For an AR(1) covariance matrix ($\Sigma_{jk}=h\rho^{|j-k|}$),  the LDL decomposition satisfies 
$l_{jk}=\rho^{|j-k|}$ if $k\leq j$, $l_{jk}=0$ if $j>k$,  $\sigma_1^2=h$, and $\sigma_k^2=h(1-\rho^2)$ for $k\geq 2$.

For a CS covariance matrix ($\Sigma_{jk}=h\rho$ if $j\neq k$, $h$ if $j=k$),  we have
$l_{jk}=1$ if $k=j$,  $\frac{\rho}{1+(k-1)\rho}$   if $k< j$, and $0$ otherwise, and $\sigma_1^2=h$, 
$\sigma_2^2=h(1-\rho^2)$, and $\sigma_{k}^2=\sigma_{k-1}^2 [1-(\frac{\rho}{1+(k-2)\rho})^2]$ for $k\geq 2$.

\bibliographystyle{wileyj}
\bibliography{sim_reml} 

\end{document}